\documentclass[10pt,nc,aps,notitlepage,superscriptaddress,twocolumn]{revtex4-2}
\usepackage{amsmath}
\usepackage{amssymb}
\usepackage{amsfonts}
\usepackage{bbm}
\usepackage{bm}
\usepackage{graphicx}
\usepackage{float}
\usepackage{dcolumn}
\usepackage{MnSymbol}
\usepackage{times}
\usepackage{verbatim}

\usepackage[utf8]{inputenc}
\usepackage[T1]{fontenc}
\usepackage[english]{babel}  

\usepackage{tikz}

\usepackage{colortbl}
\usepackage{tabu,diagbox}
\usepackage{natbib}
\setcitestyle{super}

\usepackage{times}
\usepackage{dsfont}

\usepackage{amssymb,amsmath,amsfonts,amsthm}
\usepackage{mathtools}
\usepackage{verbatim}
\usepackage{enumerate}
\usepackage{adjustbox}
\usepackage{graphicx}
\graphicspath{figs/}
\usepackage[colorlinks,linkcolor=blue,anchorcolor=blue,urlcolor=blue,citecolor=blue]{hyperref}
\usepackage[capitalise]{cleveref}
\usepackage{bbm}
\usepackage{booktabs}

\usepackage{color}
\definecolor{dred}{rgb}{.8,0.2,.2}
\definecolor{ddred}{rgb}{.8,0.5,.5}
\definecolor{dblue}{rgb}{.2,0.2,.8}
\definecolor{dgreen}{rgb}{.2,0.5,.2}

\newcommand{\bra}[1]{\mbox{$\langle #1|$}}
\newcommand{\ket}[1]{\ensuremath{|#1\rangle}}

\newcommand{\tr}{\textrm{tr}}

\newcommand{\be}{\begin{equation}}
\newcommand{\ee}{\end{equation}}
\newcommand{\bea}{\begin{eqnarray}}
\newcommand{\eea}{\end{eqnarray}}

\newcommand{\ii}{\textrm{i}}
\newcommand{\bmm}{\begin{matrix}}
\newcommand{\emm}{\end{matrix}}
\newcommand{\BLvert}{\Biggl\vert\bmm} 
\newcommand{\Brangle}{\emm\Biggr\rangle}

\makeatletter\renewcommand{\@biblabel}[1]{#1.}\makeatother

\newcommand{\fmoveh}{
\begin{tikzpicture}[x=0.75pt,y=0.75pt,yscale=-1,xscale=1]
\draw    (170,120) -- (180,140) ;
\draw    (170,160) -- (180,140) ;
\draw    (180,140) -- (200,140) ;
\draw    (210,120) -- (200,140) ;
\draw    (200,140) -- (210,160) ;
\draw (162.5,117.5) node [anchor=north west][inner sep=0.75pt]  [font=\scriptsize] [align=left] {$\displaystyle i$};
\draw (161.5,148.5) node [anchor=north west][inner sep=0.75pt]  [font=\scriptsize] [align=left] {$\displaystyle j$};
\draw (210.5,117.5) node [anchor=north west][inner sep=0.75pt]  [font=\scriptsize] [align=left] {$\displaystyle k$};
\draw (211.5,150) node [anchor=north west][inner sep=0.75pt]  [font=\scriptsize] [align=left] {$\displaystyle l$};
\draw (184,130) node [anchor=north west][inner sep=0.75pt]  [font=\scriptsize] [align=left] {$\displaystyle m$};
\end{tikzpicture}}
\newcommand{\fmovev}{
\begin{tikzpicture}[x=0.75pt,y=0.75pt,yscale=-1,xscale=1]
\draw    (279.5,119) -- (300,130) ;
\draw    (279.5,159) -- (300,150) ;
\draw    (300,130) -- (300,150) ;
\draw    (319.5,119) -- (300,130) ;
\draw    (300,150) -- (319.5,159) ;
\draw (272,116.5) node [anchor=north west][inner sep=0.75pt]  [font=\scriptsize] [align=left] {$\displaystyle i$};
\draw (271,147.5) node [anchor=north west][inner sep=0.75pt]  [font=\scriptsize] [align=left] {$\displaystyle j$};
\draw (320,116.5) node [anchor=north west][inner sep=0.75pt]  [font=\scriptsize] [align=left] {$\displaystyle k$};
\draw (321,149) node [anchor=north west][inner sep=0.75pt]  [font=\scriptsize] [align=left] {$\displaystyle l$};
\draw (302,135) node [anchor=north west][inner sep=0.75pt]  [font=\scriptsize] [align=left] {$\displaystyle n$};
\end{tikzpicture}}
\newcommand{\treebasis}{
\begin{tikzpicture}[x=0.75pt,y=0.75pt,yscale=-1,xscale=1]
\draw    (419.33,101.4) -- (454.52,150) ;
\draw    (431.06,117.6) -- (447.26,105.87) ;
\draw    (442.79,133.8) -- (458.99,122.07) ;
\draw    (414,150) -- (494,150) ;
\draw (411,92) node [anchor=north west][inner sep=0.75pt]  [font=\scriptsize] [align=left] {$\displaystyle 1$};
\draw (446.33,93.72) node [anchor=north west][inner sep=0.75pt]  [font=\scriptsize] [align=left] {$\displaystyle 1$};
\draw (458,112) node [anchor=north west][inner sep=0.75pt]  [font=\scriptsize] [align=left] {$\displaystyle 1$};
\draw (437,113.79) node [anchor=north west][inner sep=0.75pt]  [font=\scriptsize] [align=left] {$\displaystyle m$};
\draw (449.4,131.92) node [anchor=north west][inner sep=0.75pt]  [font=\scriptsize] [align=left] {$\displaystyle n$};
\draw (415,152) node [anchor=north west][inner sep=0.75pt]  [font=\scriptsize] [align=left] {$\displaystyle 1$};
\draw (495,152) node [anchor=north west][inner sep=0.75pt]  [font=\scriptsize] [align=left] {$\displaystyle 1$};
\end{tikzpicture}
}
\newcommand{\braida}{
\begin{tikzpicture}[x=0.75pt,y=0.75pt,yscale=-1,xscale=1]
\draw    (273.38,220.51) -- (296.84,252.91) ;
\draw    (285.19,236.71) -- (301.39,224.98) ;
\draw    (273.38,220.51) .. controls (262.02,204.25) and (273.24,195.31) .. (291.82,194.82) ;
\draw    (274.19,195.46) .. controls (272.38,194.21) and (267.88,191.78) .. (263.89,190.35) ;
\draw    (273.38,220.51) .. controls (283.31,213.07) and (283.95,205.58) .. (278.74,199.48) ;
\draw    (257,252.91) -- (337,252.91) ;
\draw (252,182) node [anchor=north west][inner sep=0.75pt]  [font=\scriptsize] [align=left] {$\displaystyle 1$};
\draw (294.23,185.37) node [anchor=north west][inner sep=0.75pt]  [font=\scriptsize] [align=left] {$\displaystyle 1$};
\draw (302.73,216.3) node [anchor=north west][inner sep=0.75pt]  [font=\scriptsize] [align=left] {$\displaystyle 1$};
\draw (279.73,218.97) node [anchor=north west][inner sep=0.75pt]  [font=\scriptsize] [align=left] {$\displaystyle m$};
\draw (291.07,234.63) node [anchor=north west][inner sep=0.75pt]  [font=\scriptsize] [align=left] {$\displaystyle n$};
\draw (328,254.91) node [anchor=north west][inner sep=0.75pt]  [font=\scriptsize] [align=left] {$\displaystyle 1$};
\draw (258,254.91) node [anchor=north west][inner sep=0.75pt]  [font=\scriptsize] [align=left] {$\displaystyle 1$};
\end{tikzpicture}}
\newcommand{\braidb}{
\begin{tikzpicture}[x=0.75pt,y=0.75pt,yscale=-1,xscale=1]
\draw    (124.81,101.4) -- (160,150) ;
\draw    (173,115.71) .. controls (157.65,127.13) and (153.86,104.63) .. (136.97,117.1) ;
\draw    (161.27,99.51) .. controls (156.22,103.07) and (154.91,106.59) .. (154.87,110.13) ;
\draw    (156.03,119.48) .. controls (156.3,122.65) and (155.47,128.5) .. (148.7,133.3) ;
\draw    (120,150) -- (200,150) ;
\draw (115.75,90.05) node [anchor=north west][inner sep=0.75pt]  [font=\scriptsize] [align=left] {$\displaystyle 1$};
\draw (159.5,88.8) node [anchor=north west][inner sep=0.75pt]  [font=\scriptsize] [align=left] {$\displaystyle 1$};
\draw (175,107.89) node [anchor=north west][inner sep=0.75pt]  [font=\scriptsize] [align=left] {$\displaystyle 1$};
\draw (142.75,116.05) node [anchor=north west][inner sep=0.75pt]  [font=\scriptsize] [align=left] {$\displaystyle m$};
\draw (153.75,131.05) node [anchor=north west][inner sep=0.75pt]  [font=\scriptsize] [align=left] {$\displaystyle n$};
\draw (191,152) node [anchor=north west][inner sep=0.75pt]  [font=\scriptsize] [align=left] {$\displaystyle 1$};
\draw (121,152) node [anchor=north west][inner sep=0.75pt]  [font=\scriptsize] [align=left] {$\displaystyle 1$};
\end{tikzpicture}
}
\newcommand{\subsystem}{
\begin{tikzpicture}[x=0.75pt,y=0.75pt,yscale=-1,xscale=1]
\draw    (420,130) -- (560,130) ;
\draw    (450,100) -- (450,130) ;
\draw    (490,100) -- (490,130) ;
\draw    (530,100) -- (530,130) ;
\draw  [dash pattern={on 3.75pt off 3.75pt}]  (470,130) -- (470,160) ;
\draw  [dash pattern={on 3.75pt off 3.75pt}]  (510,130) -- (510,160) ;
\draw (441,95) node [anchor=north west][inner sep=0.75pt]  [font=\scriptsize] [align=left] {$\displaystyle 1$};
\draw (481,95) node [anchor=north west][inner sep=0.75pt]  [font=\scriptsize] [align=left] {$\displaystyle 1$};
\draw (521,95) node [anchor=north west][inner sep=0.75pt]  [font=\scriptsize] [align=left] {$\displaystyle 1$};
\draw (418,132) node [anchor=north west][inner sep=0.75pt]  [font=\scriptsize] [align=left] {$\displaystyle 1$};
\draw (551,132) node [anchor=north west][inner sep=0.75pt]  [font=\scriptsize] [align=left] {$\displaystyle 1$};
\draw (461,145) node [anchor=north west][inner sep=0.75pt]  [font=\scriptsize] [align=left] {$\displaystyle 0$};
\draw (501,145) node [anchor=north west][inner sep=0.75pt]  [font=\scriptsize] [align=left] {$\displaystyle 0$};
\draw (458,115) node [anchor=north west][inner sep=0.75pt]  [font=\scriptsize] [align=left] {$\displaystyle j$};
\draw (478,115) node [anchor=north west][inner sep=0.75pt]  [font=\scriptsize] [align=left] {$\displaystyle j$};
\draw (498,115) node [anchor=north west][inner sep=0.75pt]  [font=\scriptsize] [align=left] {$\displaystyle k$};
\draw (518,115) node [anchor=north west][inner sep=0.75pt]  [font=\scriptsize] [align=left] {$\displaystyle k$};
\end{tikzpicture}}

\usepackage[small,raggedright]{titlesec}
\titleformat*{\section} {\small\bf}
\titlespacing*{\section} {0pt}{10pt}{0pt}
\titlespacing*{\subsection} {0pt}{10pt}{0pt}
\begin{document}

\title{Experimental realization of a topologically protected Hadamard gate via braiding Fibonacci anyons}

\author{Yu-ang Fan}
\affiliation{Shenzhen Institute for Quantum Science and Engineering and Department of Physics, Southern University of Science and Technology, Shenzhen 518055, China}
\author{Yingcheng Li}
\affiliation{State Key Laboratory of Surface Physics, Department of Physics, Center for Field Theory and Particle Physics, and Institute for Nanoelectronic devices and Quantum computing, Fudan University, Shanghai 200433, and Shanghai Qi Zhi Institute, Shanghai 200030, China}
\author{Yuting Hu}
\affiliation{School of Physics, Hangzhou Normal University, Hangzhou 311121, China}
\author{Yishan Li} 
\affiliation{Shenzhen Institute for Quantum Science and Engineering and Department of Physics, Southern University of Science and Technology, Shenzhen 518055, China}
\author{Xinyue Long} 
\affiliation{Shenzhen Institute for Quantum Science and Engineering and Department of Physics, Southern University of Science and Technology, Shenzhen 518055, China}
\author{Hongfeng Liu} 
\affiliation{Shenzhen Institute for Quantum Science and Engineering and Department of Physics, Southern University of Science and Technology, Shenzhen 518055, China}
\author{Xiaodong Yang} 
\affiliation{Shenzhen Institute for Quantum Science and Engineering and Department of Physics, Southern University of Science and Technology, Shenzhen 518055, China}
\author{Xinfang Nie} 
\affiliation{Shenzhen Institute for Quantum Science and Engineering and Department of Physics, Southern University of Science and Technology, Shenzhen 518055, China}
\author{Jun Li}
\affiliation{Shenzhen Institute for Quantum Science and Engineering and Department of Physics, Southern University of Science and Technology, Shenzhen 518055, China}
\author{Tao Xin}
\affiliation{Shenzhen Institute for Quantum Science and Engineering and Department of Physics, Southern University of Science and Technology, Shenzhen 518055, China}
\author{Dawei Lu}
\email{ludw@sustech.edu.cn}
\affiliation{Shenzhen Institute for Quantum Science and Engineering and Department of Physics, Southern University of Science and Technology, Shenzhen 518055, China}
\author{Yidun Wan}
\email{ydwan@fudan.edu.cn}
\affiliation{State Key Laboratory of Surface Physics, Department of Physics, Center for Field Theory and Particle Physics, and Institute for Nanoelectronic devices and Quantum computing, Fudan University, Shanghai 200433, and Shanghai Qi Zhi Institute, Shanghai 200030, China}

\def\thefootnote{*}\footnotetext{These authors contributed equally to this work}
\def\thefootnote{\arabic{footnote}}

\begin{abstract}
Topological quantum computation (TQC) is one of the most striking architectures that can realize fault-tolerant quantum computers. In TQC, the logical space and the quantum gates are topologically protected, i.e., robust against local disturbances. The topological protection, however, requires rather complicated lattice models and hard-to-manipulate dynamics; even the simplest system that can realize universal TQC--the Fibonacci anyon system--lacks a physical realization, let alone braiding the non-Abelian anyons. Here, we propose a disk model that can realize the Fibonacci anyon system, and construct the topologically protected logical spaces with the Fibonacci anyons. Via braiding the Fibonacci anyons, we can implement universal quantum gates on the logical space. Our proposal is platform-independent. As a demonstration, we implement a topological Hadamard gate on a logical qubit through a sequence of $15$ braiding operations of three Fibonacci anyons with merely $2$ nuclear spin qubits. The gate fidelity reaches 97.18\% by randomized benchmarking. We further prove by experiment that the logical space and Hadamard gate are topologically protected: local disturbances due to thermal fluctuations result in a global phase only. Our work is a proof of principle of TQC and paves the way towards fault-tolerant quantum computation.
\end{abstract}
\maketitle

Among all the schemes of quantum computation, topological quantum computation (TQC)\cite{Freedman2002} stands out because of its fault tolerance due to topological protection: In a topological quantum computer, the logical computational space is a subspace of the Hilbert space of certain number of non-Abelian anyons, whose braiding implements the logical gates; such logical gates and logical computing spaces are topologically protected against local disturbances, and are thus robust. 

Majorana fermions have been suggested to realize TQC\cite{Kitaev2001,Alicea2012,Sarma2015}, but are unsatisfactory because they cannot realize for example the Hamamard gate and hence non-universal\cite{Lahtinen2017}. The simplest non-Abelian anyons that can realize universal TQC are the Fibonacci anyons\cite{Nayak2008,Hormozi2007,Bonesteel2005}. Unfortunately, physical realizations of controllable Fibonacci anyons are still missing, let alone more complicated anyons. This conundrum is ascribed to two main reasons. On the one hand, a real material that bear Fibonacci anyons is still unknown. On the other hand, lattice models of Fibonacci anyons are complicated, and there has not been any lattice model in which all quasiparticles are Fibanocci anyons. For example, the string-net model\cite{Levin2005,Hu2018a,Hu2018,Hu2017} is a lattice model that accomadates not only Fibonacci anyons but also anti-chiral Fibonacci anyons and their composites. Realizing such models with Fibonacci anyons demands numerous degrees of freedom. The worse is, identifying and manipulating these Fibonacci anyons are beyond the reach of state-of-the-art technologies. We however finds a way out of this conundrum.

In this paper, we first propose a lattice model on the disk describing a Fibonacci anyon system, in which the boundary spectrum is chiral, i.e., only the Fibonacci anyons can be exited at the boundary of the disk and then braided to implement logical quantum gates on the logical qubits encoded in the Hilbert space of the Fibonacci anyons. Realization of our proposal is platform independent, i.e., can be done in any controllable system of physical qubits. As a demonstration of our proposal, we then implement a topological Hadamard gate on a logical qubit through a sequence of 15 braiding operations of 3 boundary Fibonacci anyons with merely 2 nuclear spin qubits. This result contrasts previous works that spends 3 qubits to realize only the ground states (no anyons) of a Fibonacci string-net model\cite{Li2017} and that spends 4 qubits to realize only the ground states of a toric code model\cite{Luo2018,Luo2019}. The gate fidelity is 97.18\% by randomized benchmarking (RB). Via purity benchmarking (PB), we found that the origin of the infidelity is the incoherent error generally caused by the dephasing of our physical system. 

A working topological quantum computer may suffer local disturbances due to thermal fluctuations. A topological quantum computer works at controlled, extremely low temperature, at which the thermal fluctuations cannot produce any real Fibonacci anyons on top of the existing Fibonacci anyons used for computation. Then the thermal fluctuations can only produce paired Fibonacci anyons, which may interfere with the braiding of the nearby real Fibonacci anyons. Fortunately, the logical space and gates have been argued not to be affected at all by such disturbances\cite{Nayak2008}, viz topologically protected. In this paper, we prove the topological protection by experiment: local disturbances due to thermal fluctuations result in a global phase only.

By realizing the topological Hadamard gate on three Fibonacci anyons and demonstrating the topological protection, our work is a proof of principle of Fibonacci-anyon based TQC.   


\begin{figure*}
    \centering
    \includegraphics[width=0.77\linewidth]{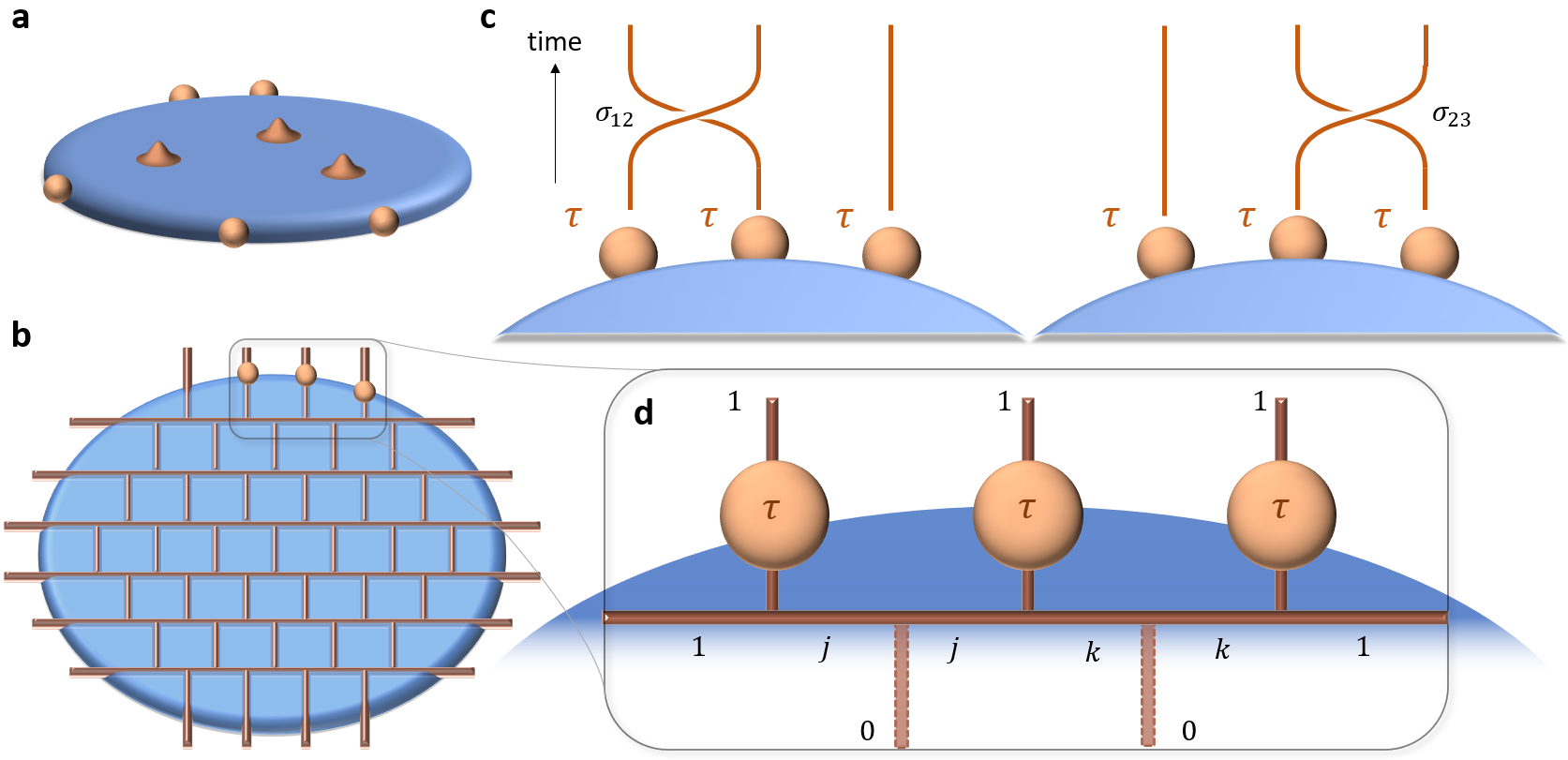}
    \caption{\textbf{Fibonacci anyon system and the model.} \textbf{a}, Doubled Fibonacci topological phase with a gapped boundary. The bumps on the disk are possible bulk excitations but irrelevant. The balls on the boundary depict boundary Fibonacci anyons. \textbf{b}, Trivalent lattice that discretizes the disk. Boundary Fibonacci anyons (orange balls) reside on open edges. \textbf{c}, Three boundary Fibonacci anyons encoding the logical qubit. The braid generator $\sigma_{12}$ ($\sigma_{23}$) braids the worldlines of the first and the second (the second and the third) Fibonacci anyons. \textbf{d}, Minimal subsystem (conisiting of two degrees of freedom $j$ and $k$) that bears three boundary Fibonacci anyons.}
    \label{subsystem}
\end{figure*}

\section*{A Fibonacci anyon system and the logical qubit}

Our model describes the doubled Fibonacci topological order with a gapped boundary. Along the gapped boundary, there can be Fibonacci anyons (Fig. \ref{subsystem}a). The model is defined on a disk, discretized by a trivalent lattice, as shown in Fig. \ref{subsystem}b. The open edges and the internal edges respectively carry the boundary and bulk degrees of freedom, which take value in $\{0,1\}$. The Hilbert space of the model is spanned by all possible configurations of the degrees of freedom on the edges. An open edge taking value $1$ bears a boundary Fibonacci anyon. We can braid the boundary Fibonacci anyons by moving them around each other through the bulk. Braiding the Fibonacci anyons can be represented by certain matrices acting on the Hilbert space of the model.

According to Ref.\cite{Nayak2008}, we need in our model three neighbouring boundary Fibonacci anyons to construct a logical qubit. We can braid the three Fibonacci anyons to implement the single-qubit logical gates. All possible braiding operations of the three Fibonacci anyons can be generated by two basic operations, $\sigma_{12}$ and $\sigma_{23}$ (see Fig. \ref{subsystem}c), and their inverses. The operation $\sigma_{12}$ ($\sigma_{23}$) exchanges the left (right) two Fibonacci anyons' worldline on top. See the Seupplemental Information for more details of the model.

As a subspace of the Hilbert space of our model, the logical qubit is $2$-dimensional and is invariant under braiding the three boundary Fibonacci anyons. To operate on the logical qubit, we need to express the braid generators $\sigma_{12}$ and $\sigma_{23}$ explicitly as matrices, which depend on the positions of the three Fibonacci anyons on the boundary. In order to realize a logical qubit with the smallest number of physical qubits, we bipartite the system into two disentangled parts: a minimal subsystem (shown in Fig. \ref{subsystem}b) that bears three boundary Fibonacci anyons, supporting a logical qubit, and the subsystem consisting the rest degrees of freedom of the total system. The minimal subsystem has only two degrees of freedom $j$ and $k$. In this setting, any braiding operation of the three boundary Fibonacci anyons decomposes into the tensor product of a $4\times 4$ matrix acting on $\ket{j,k}$ and an identity matrix on the other subsystem.

As such, the logical space in the basis of $\ket{j,k}$ is spanned by $\ket{0}_{\textrm{L}}=\frac{1}{\phi}\ket{01}+\sqrt{\frac{1}{\phi}}\ket{11}$, and $\quad\ket{1}_{\textrm{L}}=-\frac{1}{\phi^{3/2}}\ket{01}+\sqrt{\frac{1}{\phi}}\ket{10}+\frac{1}{\phi^2}\ket{11}$, where $\phi=\frac{1+\sqrt{5}}{2}$ is the golden ratio.
See Methods for the two braiding generators $\sigma_{12}$ and $\sigma_{23}$. Details of constructing the logical qubit can be found in the Supplemental Information. One can realize any single-qubit gate on the logical qubit with arbitrary high precision by sequentially implementing the $\sigma_{12}$ and $\sigma_{23}$. As an example, a Hadamard gate can be realized by implementing the following sequence of braiding operations\cite{Rouabah2021}
\begin{align}
&H_\text{L}=(\sigma_{12})^{4}(\sigma_{23})^{-2}(\sigma_{12})^{2}(\sigma_{23})^{-2}(\sigma_{12})^{2}(\sigma_{23})^{2}\times\notag
\\&(\sigma_{12})^{-2}(\sigma_{23})^{4}(\sigma_{12})^{2}(\sigma_{23})^{-2}(\sigma_{12})^{-2}(\sigma_{23})^{2}(\sigma_{12})^{2}.
\label{H_L}
\end{align}

To realize a single-qubit topological quantum computer, we shall simulate the subsystem in Fig. \ref{subsystem}b and the logical Hadamard gate Eq. \eqref{H_L}. Using our model, it is straightforward to scale up the topological quantum computer by considering larger subsystems.

\section*{Experimental implementation of the Hadamard gate} 
\begin{figure*}
\centering
\includegraphics[width=1\linewidth]{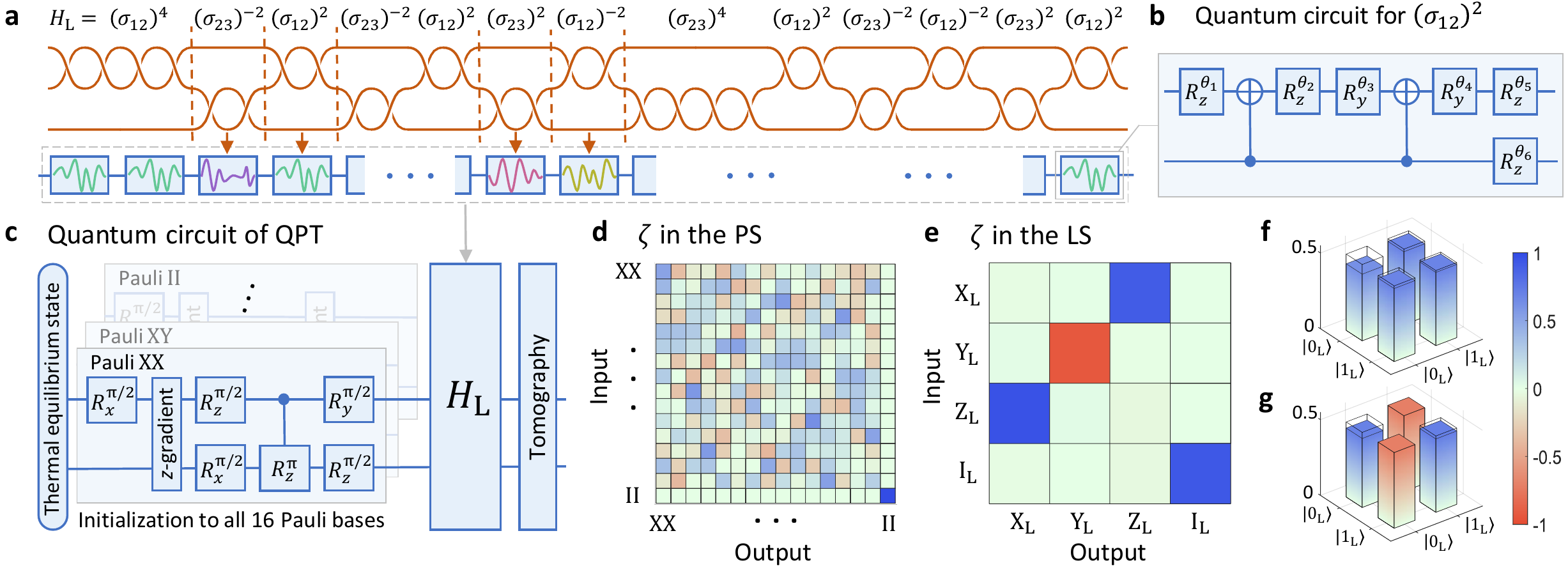}
\caption{\textbf{Implementation and process tomography of the Hadamard gate on the logical qubit.} \textbf{a}, Logical Hadamard gate $H_\text{L}$ is a sequence of 15 braiding operations, each realized by several basic quantum gates about 2 ms in total. \textbf{b}, Quantum circuit to realize $(\sigma_{12})^2$ with controlled-NOT gates and single-qubit rotations; see Methods. \textbf{c}, Quantum circuit of QPT, which requires 16 independent experiments with each Pauli basis in the PS as the input. Here, we show the circuit to prepare XX that includes single-qubit rotations and a controlled-phase gate from the thermal equilibrium state. The $z$ gradient is a gradient field along the $z$-axis to destroy coherence terms. The final tomography is a full state tomography in the PS. \textbf{d}, QPT results in in the PS. The $\zeta$ matrix of $H_\text{L}$ sets a mapping between each of the 16 Pauli inputs and outputs. \textbf{e}, QPT results in the LS. Ideally, a Hadamard gate will transform Pauli matrices in the following order: X $\rightarrow$ Z, Y $\rightarrow$ -Y, and  Z $\rightarrow$ X. The experimental result matches very well with this prediction. \textbf{f}, Density matrix in the LS when applying the $H_\text{L}$ to $|0_\text{L}\rangle$, where the theoretical output is $(|0_\text{L}\rangle+|1_\text{L}\rangle)/\sqrt{2}$. Transparent and colored bars represent the theoretical and experimental values. \textbf{g}, Density matrix in the LS when applying the $H_\text{L}$ to $|1_\text{L}\rangle$, where the theoretical output is  $(|0_\text{L}\rangle-|1_\text{L}\rangle)/\sqrt{2}$.}
\label{pt}
\end{figure*}

We use the nuclear magnetic resonance (NMR) quantum register $^{13}\text{C}$-labeled chloroform to construct the logical Hadamard gate\cite{xin2020quantum}. The $^{13}\text{C}$ and $^{1}\text{H}$ spins serve as two physical qubits, each being
controlled by a radio-frequency (rf) field \cite{nie2020experimental,lin2022experimental}. The total Hamiltonian is \cite{vandersypen2005nmr}
\begin{equation}
 \hat{\mathcal{H}}_{\text{NMR}} = \frac{\pi J}{2} \hat{\sigma}^1_z\hat{\sigma}^2_z+\sum_{i=1}^2\pi B_i(\cos\phi_i\hat{\sigma}^i_x+\sin\phi_i\hat{\sigma}^i_y),
 \label{Hnmr}
 \end{equation}
where $J =215$ Hz is the coupling strength between the qubits, $B_i$ and $\phi_i$ are tunable amplitude and phase of the $i$-th rf fields, and $\hat{\sigma}_{x,y,z}^i$ are the Pauli matrices acting on qubit $i$. All experiments are carried out on a Bruker AVANCE 600 MHz spectrometer at 295 K.

With the two physical qubits, we first decompose each of the four braiding operations: $\sigma_{12}^{+2}$, $\sigma_{12}^{-2}$, $\sigma_{23}^{+2}$ and $\sigma_{23}^{-2}$ into a quantum circuit of controlled-NOT gates and single-qubit rotations, which is further optimized by a 2 ms shaped pulse as shown in Figs. \ref{pt}a and \ref{pt}b. All of these building-block pulses have a simulated fidelity higher than 0.998, and a 15-step concatenation of them in the order as they appear in Eq. \eqref{H_L} to construct the Hadamard gate $H_\text{L}$. Hence, the experimental length of $H_\text{L}$ is 30 ms, leading to non-negligible decoherence error. Prior to experiment, we simulate the loss of fidelity by accounting for the dephasing effect caused by $T_2$ (${T_2}^\ast$) of the NMR processor being used, and find that the final fidelity of $H_\text{L}$ will drop to 98.23\% (94.63\%); see Methods. These two values roughly provide the upper- and lower-bound of the experimental fidelity of $H_\text{L}$ for the following reasnon. On the one hand, the implementation of $H_\text{L}$ will to some extent dynamically decouple the physical qubits from the environment (thus prolong ${T_2}^\ast$). On the other hand, the experimental results cannot exceed the limitation by the intrinsic $T_2$. As to be seen, the results will indeed fall within the above predicted fidelity region. 

Quantum process tomography (QPT) is the most straightforward technique to evaluate a certain quantum process. It has the advantage to uncover the complete information of the process. In experiment, we start from each of the 16 Pauli bases (XX, XY, ... , II) in the 2-qubit PS, apply the $H_\text{L}$ gate in a sequence of the braiding operations in Eq. (\ref{H_L}), and perform full state tomography on each corresponding final state; See Fig. \ref{pt}c for the quantum circuit. The state tomography in NMR is implemented in the two-qubit Pauli bases \cite{zhang2022identifying,nie2022experimental}, meaninmplete reconstruction of $H_\text{L}$ in the physical space (PS) is described by a 16-by-16 superoperator $\zeta$. The tomographic result of $\zeta$ in the PS is shown in Fig. \ref{pt}d. In addition, we transform the result in the PS to the $\zeta$ superoperator in the logical space (LS), which is more straightforward to visualize the performance of a Hadamard gate that maps X $\rightarrow$ Z, Y $\rightarrow$ -Y, and  Z $\rightarrow$ X; see Fig. \ref{pt}e.

We employ the average gate fidelity, defined as $\bar{F}=\int d\psi\langle\psi|H^{\dagger}\zeta(\psi)H|\psi\rangle$ to evaluate the performance of $\zeta$. This formula can be further simplified to enable direct fidelity computation when $\zeta$ has been fully reconstructed \cite{nielsen2002simple}. In our experiment, the average gate fidelity of $H_\text{L}$ by QPT is $F_{\text{QPT}}^{\text{PS}}=95.07\%$ and $F_{\text{QPT}}^{\text{LS}}=96.86\%$ in the physical and logical spaces. The results fall into the predicted fidelity region, but the result in the LS is slightly higher than that in the PS. In addition, for better visualization, we apply the gate on the fiducial states $|0_\text{L}\rangle$ and $|1_\text{L}\rangle$, and depict in Figs. \ref{pt}f and \ref{pt}g the output density matrices, which are supposed to be simply $(|0_\text{L}\rangle+|1_\text{L}\rangle)/{\sqrt{2}}$ and $(|0_\text{L}\rangle-|1_\text{L}\rangle)/{\sqrt{2}}$.

\begin{figure*}
\centering
\includegraphics[width=1\linewidth]{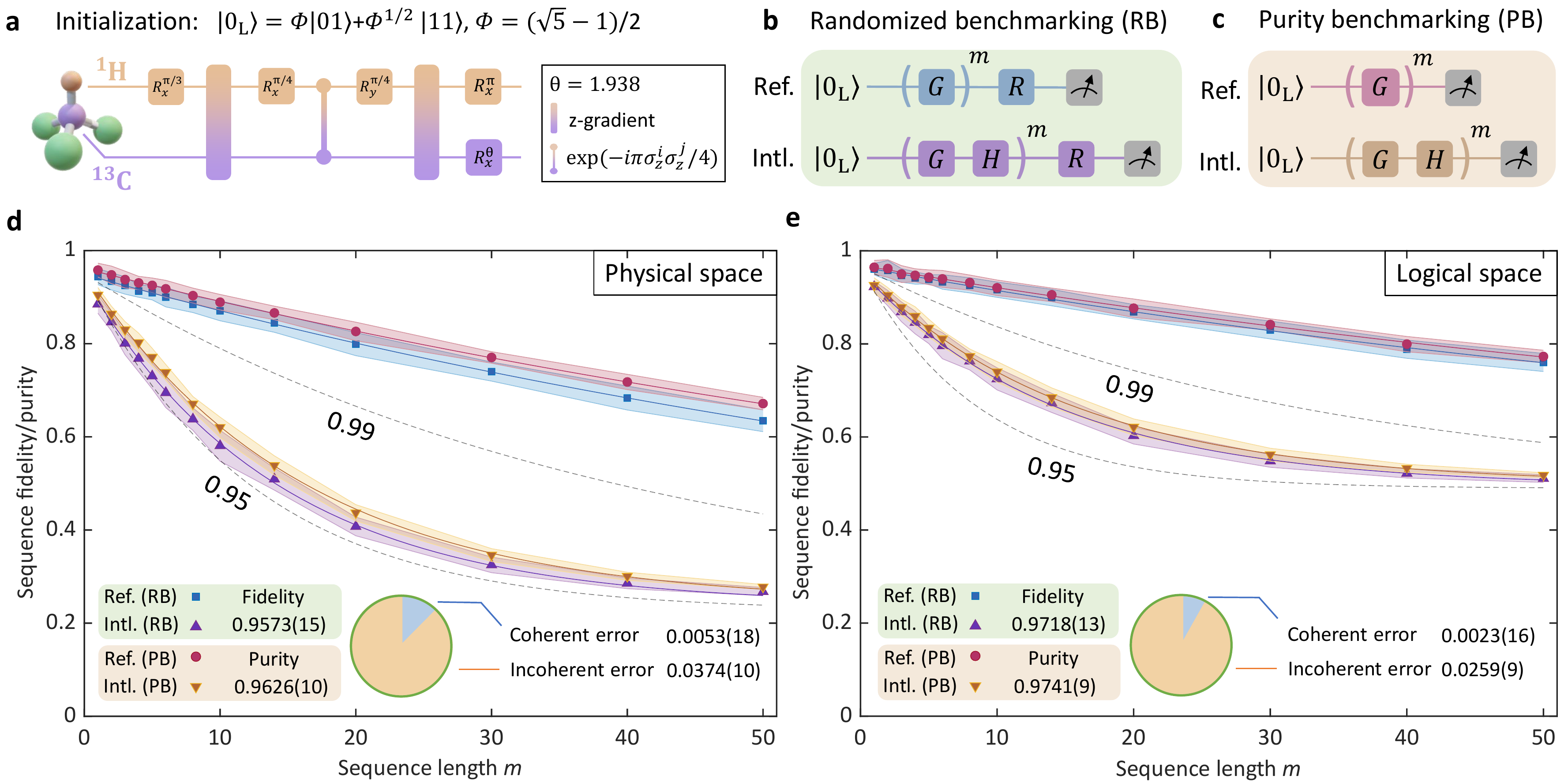}
\caption{\textbf{RB and PB results of the logical Hadamard gate.} \textbf{a}, Quantum circuit to initialize the physical qubits from the thermal equilibrium state to $|0_\text{L}\rangle$. This state preparation stage takes 4 ms. \textbf{b}, Reference (upper panel) and interleaved (lower panel) RB sequences. A reference experiment is implemented by a sequence of $m$ random Cliffords and a recovery gate $R$. To estimate the RB fidelity of $H_\text{L}$, the gate is interleaved with $m$ random Cliffords. The readout is to measure the residual $|0_\text{L}\rangle$ with respect to the initial. The difference between the reference and interleaved decay gives the RB fidelity of $H_\text{L}$. \textbf{c}, Reference (upper panel) and interleaved (lower panel) PB sequences. Compared with the RB sequence, the recovery gate is removed in PB, and the measurement is to extract purity by state tomography in the PS. \textbf{d}, Results in the PS, including the fidelity/purity decay curves as a function of sequence length $m$ for RB and PB. The fidelity/purity for each $m$ is measured for 30 different sequences, leading to error bars that originate from standard deviations. The dashed lines indicate the thresholds for exceeding gate fidelities of 0.95 and 0.99. \textbf{e}, Results in the LS, where the parameter settings are the same as those in the PS.}
\label{rb}
\end{figure*}

Despite a full description of the quantum channel, QPT inherently incorporates the state preparation and measurement (SPAM) errors and thus is not an accurate metric that completely accounts for the gate infidelity. To solve this issue, we apply the Clifford-based RB \cite{knill2008randomized,magesan2012efficient,chow2009randomized,barends2014superconducting}, where the $H_\text{L}$ gate is characterized by interleaving it with many random sequences of gates, such that the measured fidelity is resilient to the SPAM errors. The RB experiments start from $\left|0_\text{L}\right\rangle$, which can be prepared in the PS with the circuit in Fig. \ref{rb}a. The RB sequence in the LS is shown in Fig. \ref{rb}b, where $G$ is a Clifford gate randomly chosen from the single-qubit Clifford group (Supplemental Information), and $R$ is a recovery Clifford that inverts the sequence. A reference RB (upper panel in Fig. \ref{rb}b) that includes $m$ Cliffords over $k=30$ random sequences is first implemented, and the reference sequence fidelity $F_{\text{seq}}$ is fitted to $F_{\text{seq}}=A+Bf^m$. Here, $f$ is the sequence decay rate, and parameters $A$ and $B$ capture the SPAM errors. The average fidelity per reference Clifford, denoted by $F_{\text{ref}}$, can be estimated by $1-F_{\text{ref}} = (1-f)(d-1)/d$, where $d=2^N$ is the dimension of the relevant Hilbert space. Subsequently, the RB fidelity $F_\text{RB}$ of $H_\text{L}$ can be measured by interleaving it with $m$ random Cliffords (lower panel in Fig. \ref{rb}b). The equation is $1-F_{\text{RB}}=(1-f_{\text{int}}/f)(d-1)/d$, where $f_{\text{int}}$ is the interleaved sequence decay rate.

The RB results for $H_\text{L}$ in the PS and LS are illustrated in Figs. \ref{rb}d and \ref{rb}e. All $G$ and $R$ gates are generated in the PS by 5-ms shaped pulses with a simulated fidelity over 0.999. In the PS, the average fidelity per reference Clifford is $99.15\%$, and the fidelity of $H_\text{L}$ is $F_\text{RB}^\text{PS} = 95.73\%$, which is located within the predicted fidelity region and shows 0.66\% improvement compared to the QPT result. In the LS, the reference fidelity is $99.44\%$, and $F_\text{RB}^\text{LS} = 97.18\%$ (0.47\% higher than the QPT result). Although the improvement seems not prominent compared to the previous RB experiments \cite{chow2009randomized,corcoles2013process}, this result is indeed as expected qualitatively. In the previous experiments, the SPAM is a dominate error source that contributes non-negligibly to the total gate infidelity because its length and complexity are often comparable to the target Clifford gate. Nevertheless, in our experiment, the target gate $H_\text{L}$ is executed by a 15-step braiding, 30-ms shaped pulse, which is much longer than the SPAM that only takes about 4 ms. In other words, the SPAM error is merely a minor factor to the total infidelity of $H_\text{L}$, so RB's enhancement, which relies on the erasure of SPAM errors, is not as obvious as usual.   

To support our claim and further quantify the coherent and incoherent noises, we implement another group of experiments using PB \cite{wallman2015estimating,feng2016estimating}. In addition to distinguishing the coherent and incoherent errors, the PB results also provide information about how to reduce the errors, by, e.g.,  improving calibration or engineering robust pulses. Purity benchmarking can distinguish the coherent and incoherent errors because it is based on a coherent noise process.  Such a noise process is usually due to systematic control errors and can be corrected by directly reversing the unitary with perfect control, and hence does not decrease the state purity $p=\text{tr}(\rho^2)$. In contrast, the incoherence noise such as $T_1$ and $T_2$ processes will eventually minimize the purity. Therefore, a PB sequence can be designed as shown in Fig. \ref{rb}c, where the recovery gate is removed and the readout is to extract the purity via state tomography. This sequence purity decay can be used to analyze the incoherent error of the target gate, such that the coherent error can be consequently computed in association with the RB results (Supplemental Information).  

The PB results for the corresponding spaces are shown in Figs. \ref{rb}d and \ref{rb}e. In the PS, the gate infidelity of $H_\text{L}$ is $1-F_\text{RB}^\text{PS} = 4.27\%$, which consists of $3.74\%$ coherent and $0.53\%$ incoherent errors; see the pie chart in Fig. \ref{rb}d. The case in the LS is similar: The gate infidelity is $1-F_\text{RB}^\text{LS} = 2.82\%$ which is comprised of $2.59\%$ coherent and $0.23\%$ incoherent errors; see the pie chart in Fig. \ref{rb}e. The data clearly tell that the incoherent error, mainly attributed to the $T_2$ process, is the dominant error source in $H_\text{L}$. This conclusion can also be visualized from the decay curves, where both the reference and the interleaved RB and PB decay curves of $H_\text{L}$ are close to each other in both spaces. 

\section*{Test of robustness}

\begin{figure}[ht]
    \centering
    \includegraphics[width=\linewidth]{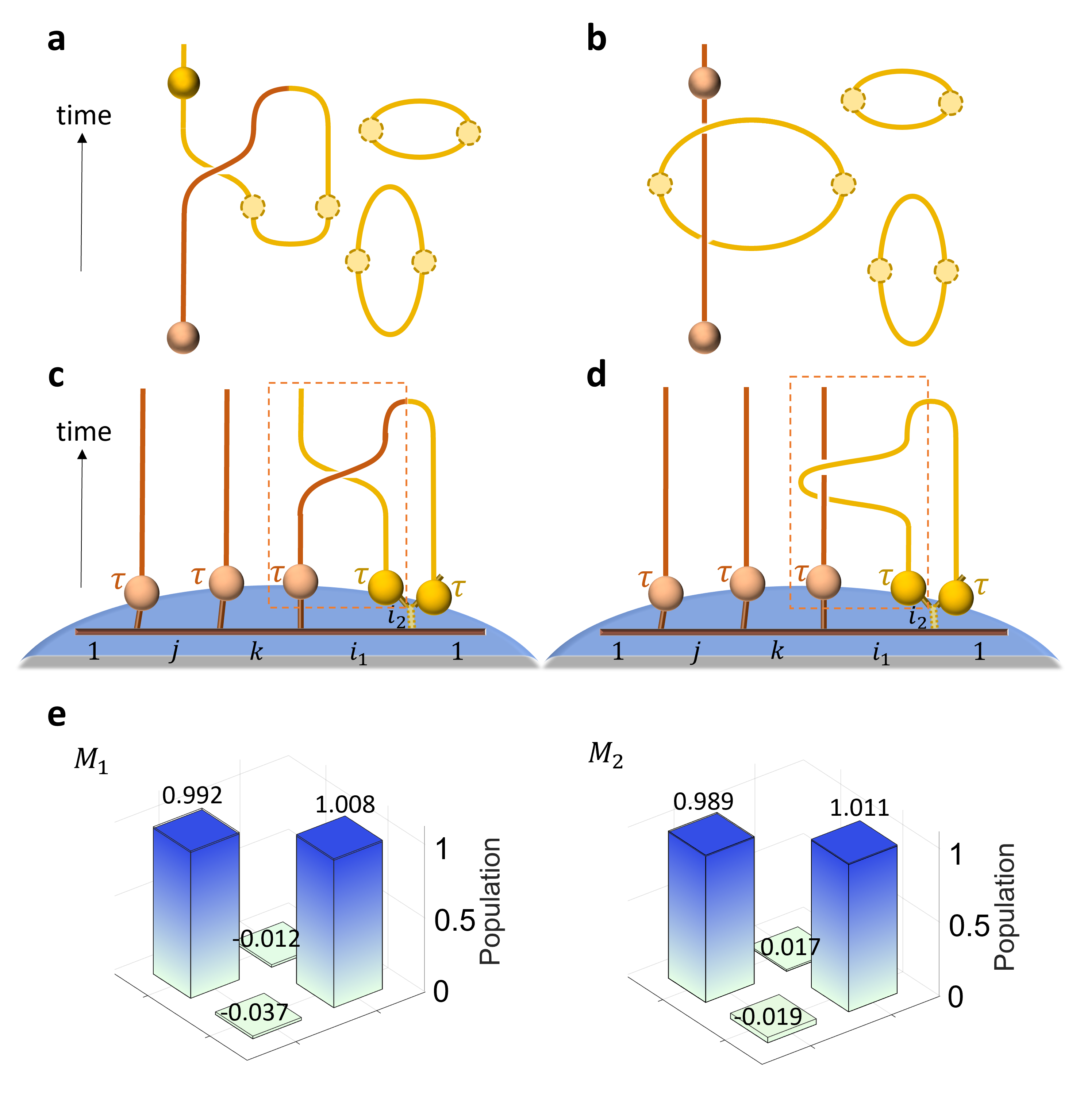}
    \caption{\textbf{Two scenarios of potential source of error in a topological quantum computer.} \textbf{a}, Scenario 1: A pair of Fibonacci anyons created from the vacuum; One in the pair may annihilate with a tracked Fibonacci anyon, while the other replaces the tracked anyon. \textbf{b}, Scenario 2: A pair of Fibonacci anyons created from the vacuum; One in the pair may braid with a tracked Fibonacci anyon and then annihilate with the other in the pair. \textbf{c} and \textbf{d}, Realizing the two scenarios in our model. \textbf{e}, $M_1$ and $M_2$ from our experiment, which are supposed to be the identity matrices in the ideal case.}
    \label{fig:anyonrobust}
\end{figure}

At low temperature, the lowest excitations---Fibonacci anyon pair states are the leading contribution to the thermal fluctuation of the system. Consequently, in a realistic topological quantum computer, the main, potential source of error are these thermally created anyon pairs. These thermally created anyons, moving around before they annihilate, may braid with the tracked anyons, i.e., the anyons for computation. We may keep the tracked anyons sufficiently far apart such that one created anyon can only braid with at most a single tracked anyon. Therefore, it suffices to consider only two scenarios \cite{Nayak2008}:
1. A pair of Fibonacci anyons are created from the vacuum. One in the pair may annihilate with a tracked Fibonacci anyon, as shown in Fig. \ref{fig:anyonrobust}a. 2. A pair of Fibonacci anyons are created from the vacuum. One in the pair may braid with a tracked Fibonacci anyon and then annihilate with the other in the pair, as shown in Fig. \ref{fig:anyonrobust}b.

Though the braiding processes of the tracked anyons are interfered in both scenarios, it is argued that no error would occur in either scenario \cite{Nayak2008}. We shall show this robustness by experiment, as a proof of principle.

To replicate in our system the two scenarios in Figs. \ref{fig:anyonrobust}a and \ref{fig:anyonrobust}b, we consider two boundary Fibonacci anyons created beside our subsystem, as shown in Figs. \ref{fig:anyonrobust}c and \ref{fig:anyonrobust}d. Two more degrees of freedom $i_1$ and $i_2$ arise, in addition to the existing degrees of freedom $j$ and $k$. So as to keep the two new Fibonacci anyons paired with each other, at the beginning and the end of the scenarios, we demand that $i_1=1$ and $i_2=0$. In scenario 1 (2), a tracked Fibonacci anyon and one of the thermally created Fibonaaci anyons are braided once (twice), as shown in the orange dashed box in Figs. \ref{fig:anyonrobust}c and \ref{fig:anyonrobust}d.

We denote scenarios 1 and 2 by two operators $\Gamma_1$ and $\Gamma_2$. We expect that, after implementing $\Gamma_1$ or $\Gamma_2$, the state of the logical qubit differs only by a global phase factor. Namely,
\begin{align}\label{robustness}
\Gamma_{q} (a\ket{0}_{\textrm{L}}+b\ket{1}_{\textrm{L}})\ket{10}_{\textrm{E}}
=e^{\ii \theta_q}(a\ket{0}_{\textrm{L}}+b\ket{1}_{\textrm{L}})\ket{10}_{\textrm{E}},
\end{align}
where $q=1, 2$ denotes that the tracked and thermally created Fibonaaci anyons are braided once or twice, and $\ket{10}_{\textrm{E}}$ denotes $\ket{i_1=1,i_2=0}$. The information stored in the logical qubit is hence intact, so are any subsequent gate operations. We verify Eq. \eqref{robustness} by verifying its sufficient and necessary condition:
\begin{align}\label{condition}
    (M_q)_{ij}\equiv{_{\textrm{L}}\hspace{-1pt}\bra{i}{_{\textrm{E}}\hspace{-1pt}\bra{10}} \mathcal{B}_q \ket{j}_{\textrm{L}}\ket{10}_{\textrm{E}}}\propto\left(\begin{array}{cc}1 & 0\\0&1\end{array}\right),
\end{align}
where $\mathcal{B}_q$ braids a tracked Fibonacci anyon and a thermally created one $q$ times. See the Supplemental Information for a derivation of condition in Eq. \eqref{condition}.

We use a 4-qubit NMR sample crotonic acid to experimentally prove the robustness of the logical qubit against the interference of thermally created anyon pairs, i.e., verifying Eq. \eqref{condition}. The system is initialized in the input states $\ket{0}_{\textrm{L}}\ket{10}_{\textrm{E}}$ and $\ket{1}_{\textrm{L}}\ket{10}_{\textrm{E}}$ with fidelity over 0.99. Two optimized 48-ms control pulses are used to implement the $\mathcal{B}_1$ and $\mathcal{B}_2$. Via full state tomography conducted in the PS and maximum likelihood estimation, we find the two matrices $M_1$ and $M_2$ in Eq. \eqref{condition} from our experiment, depicted in Fig. \ref{fig:anyonrobust}e. The two figures indicate that the conditions are well satisfied. Hence, it is proved that the logical qubit is uninfluenced in both scenarios.

\section*{Conclusion}

We have proposed a disk model and constructed topologically protected logical spaces by the boundary Fibonacci anyons. We have implemented a topological Hadamard gate on a logical qubit via a sequence of 15 braiding operations. We further have proved by experiment that the logical space and Hadamard gate are topologically protected. Further developments should include the realization of a topological entangling gate on two logical quibts, but such gates are plagued by leakage in principle. So far, it remains unknown how to realize an entangling gate without leakage\cite{Ainsworth2011}. Arbitrarily small leakage may however be achieved by tremendously more braiding operations\cite{Cui2019,Xu2008}, causing much longer gate operation time.

\section*{Methods}

\textbf{Constructing the logical qubit}
To construct the logical qubit, we need a basis transformation called $F$ move,
\begin{equation*}
\BLvert\fmoveh\Brangle=\sum_n F^{ijm}_{kln}\BLvert\fmovev\Brangle.
\end{equation*}
Here, the coefficient $F^{ijm}_{kln}$ is called a $6j$ symbol. We can then transform our original basis of the subsystem in the main text into a new `tree' basis where $\sigma_{1}$ and $\sigma_{2}$ are block-diagonalized,
\begin{align*}
    &\BLvert\scalebox{0.8}{\subsystem}\Brangle=\\
    &\sum_{mn}F^{11j}_{1km}F^{11k}_{m1n}\BLvert\scalebox{0.8}{\treebasis}\Brangle
\end{align*}

See the supplemental material for details of the $6j$ symbols. Here we only give the numerical value of the basis transformation matrix,
\begin{align*}
&[F]_{\{m,n\},\{i,j\}}\equiv F^{11j}_{1km}F^{11k}_{m1n}=\\
&\left(
\begin{matrix}
1&0&0&0\\
0&\frac{2}{1+\sqrt{5}}&0&\sqrt{\frac{2}{1+\sqrt{5}}}\\ 
0 & \frac{2}{1+\sqrt{5}} & \frac{2}{1+\sqrt{5}} & -\frac{2\sqrt{2}}{(1+\sqrt{5})^{3/2}}\\
0& -\frac{2\sqrt{2}}{(1+\sqrt{5})^{3/2}}&\sqrt{\frac{2}{1+\sqrt{5}}}&(\frac{2}{1+\sqrt{5}})^2
\end{matrix}\right);
\end{align*}

The braid generators $\sigma_{12}$ and $\sigma_{23}$ in the new `tree' basis $\ket{mn}_{\textrm{tree}}$ are represented as

\begin{align*}
&\sigma_{12}\BLvert\scalebox{0.85}{\treebasis}\Brangle=\BLvert\scalebox{0.85}{\braida}\Brangle\\
&=\sum_{m',n'}[B_1]_{\{mn\},\{m'n'\}}\BLvert\scalebox{0.7}{\treebasis{i}}\Brangle
\end{align*}

and

\begin{align*}
&\sigma_{23}\BLvert\scalebox{0.85}{\treebasis{j}}\Brangle=\BLvert\scalebox{0.85}{\braidb}\Brangle\\
&=\sum_{m',n'}[B_2]_{\{mn\},\{m'n'\}}\BLvert\scalebox{0.7}{\treebasis{i}}\Brangle,
\end{align*}
where $B_1$ and $B_2$ are given by
\begin{align*}
&B_1=\left(
\bmm
1&0&0&0\\
0&e^{-\ii \frac{4\pi}{5}}&0&0\\
0&0&e^{\ii \frac{3\pi}{5}}&0\\
0&0&0&e^{\ii \frac{3\pi}{5}}\\
\emm
\right)
\\
&B_2=\left(
\bmm
1 & 0 & 0 & 0\\
0 & e^{\ii \frac{3\pi}{5}}\frac{2}{1+\sqrt{5}} & 0 & -\ii e^{-\ii \frac{\pi}{10}} \sqrt{\frac{2}{1+\sqrt{5}}}\\
0 & 0 &  e^{\ii \frac{3\pi}{5}}&0\\
0 & -\ii e^{-\ii \frac{\pi}{10}} \sqrt{\frac{2}{1+\sqrt{5}}} &0 & -\frac{2}{1+\sqrt{5}}\\
\emm
\right).
\end{align*}
Now we see that $\ket{01}_{\textrm{tree}}$ and $\ket{11}_{\textrm{tree}}$ span a $2$-dimensional subspace that is invariant under $\sigma_{12}$ and $\sigma_{23}$. One then recovers $\ket{0}_{\textrm{L}}$ and $\ket{1}_{\textrm{L}}$ in the main text by letting $\ket{0}_{\textrm{L}}=\ket{01}_{\textrm{tree}}$ and $\ket{1}_{\textrm{L}}=\ket{11}_{\textrm{tree}}$. 

\textbf{Braiding operations and their realizations.}
The braiding generators $\sigma_{12}$ and $\sigma_{23}$ are represented in the basis $\ket{j,k}$ as
\begin{align*}
&\sigma_{12}=\left(
\bmm
1 & 0 & 0 & 0\\
0 & e^{\ii \frac{3\pi}{5}}\frac{2}{1+\sqrt{5}} & 0 & e^{\ii \frac{7\pi}{5}} \sqrt{\frac{2}{1+\sqrt{5}}}\\
0 & 0 & e^{\ii \frac{3\pi}{5}} & 0\\
0 & e^{\ii \frac{7\pi}{5}} \sqrt{\frac{2}{1+\sqrt{5}}} & 0 & -\frac{2}{1+\sqrt{5}}
\emm
\right),\\
&\sigma_{23}=\left(
\bmm
1 & 0 & 0 & 0\\
0 & e^{\ii \frac{3\pi}{5}} & 0 & 0\\
0 & 0 & e^{\ii \frac{3\pi}{5}}\frac{2}{1+\sqrt{5}} & e^{\ii \frac{7\pi}{5}} \sqrt{\frac{2}{1+\sqrt{5}}}\\
0 & 0 & e^{\ii \frac{7\pi}{5}} \sqrt{\frac{2}{1+\sqrt{5}}} & -\frac{2}{1+\sqrt{5}}
\emm
\right).
\end{align*}
In the LS, the braiding generators $\sigma_{12}$ and $\sigma_{23}$ are represented by
\begin{align}
&(\sigma_{12})_{\textrm{L}}=\left(
\bmm
e^{-\ii \frac{4\pi}{5}}&0\\
0&e^{\ii \frac{3\pi}{5}}\\
\emm
\right),\notag
\\
&(\sigma_{23})_{\textrm{L}}=\left(
\bmm
e^{\ii \frac{3\pi}{5}}\frac{2}{1+\sqrt{5}}  & e^{\ii \frac{7\pi}{5}} \sqrt{\frac{2}{1+\sqrt{5}}}\\
e^{\ii \frac{7\pi}{5}} \sqrt{\frac{2}{1+\sqrt{5}}} & -\frac{2}{1+\sqrt{5}}
\emm
\right).\notag
\end{align}

To realize these braiding operations, we decompose them into two controlled-NOT gates and several single-qubit rotations. Taking the $(\sigma_{12})^{2}$ in Fig. 2b as an example, we choose the parameters as $\theta_1=0.314, \theta_2=-0.628, \theta_3=-1.179, \theta_4=1.179, \theta_5=-2.1991$, and $\theta_6=1.885$. For the operation $(\sigma_{12})^{-2}$, the circuit is the same while the parameters are $\theta_1=2.827, \theta_2=-2.513, \theta_3=-1.179, \theta_4=1.179, \theta_5=2.1991$, and $\theta_6=2.827$. For the operations of $(\sigma_{23})^{2}$ and $(\sigma_{23})^{-2}$, we just need to swap the two qubits in the quantum circuit for $\sigma_{12}$ with the same parameters as those for $(\sigma_{12})^{2}$ and $(\sigma_{12})^{-2}$ because the forms of $\sigma_{12}$ and $\sigma_{23}$ only differ by a swap gate. 

\textbf{Numerical simulation of decoherence.} To simulate the decoherence numerically, we make the following assumptions: 
the environment is Markovian; the system and the environment are uncorrelated at the beginning; the dephasing error on each qubit is independent of that on any other qubit; the imperfections of the shaped pulses are taken into account. 

To solve the master equation, we assume that the dissipator $D$ and the total Hamiltonian $H_{tot}$ commute for short times. Therefore, the evolution of the state was simulated in two steps: evolution by $-D e^{iH_{tot}\Delta t}$, and subsequent dephasing for $\Delta t$, which was chosen to match the pulse discretization. The dephasing channel implements the exponential decay of the off-diagonal elements according to the relevant linear combinations of $T_2$ values of $^{13}$C and $^{1}\text{H}$. 

\textbf{Purity benchmarking.} We assume that the total gate error can be characterized by a completely-positive-trace-preserving linear map $\varepsilon:\mathcal{B}(\mathbb{C}^d)\rightarrow \mathcal{B}(\mathbb{C}^d)$. The coherence of a noisy process can be quantified by the unitarity of the corresponding mapping. To define the unitarity, one can use the average purity of output states over all pure states. Nevertheless, by this definition, the non-unitary channel $\varepsilon_0(\psi)=\tr(\psi)|0\rangle\langle0|$ share the same unitarity with a unitary channel. An improved idea is to subtract the identity components\cite{wallman2015estimating}, which composes another channel $\varepsilon^\prime(\psi)=\varepsilon(\psi)-\frac{1}{\sqrt{d}}\tr(\varepsilon(\psi))\mathbb{I_d}$. The unitarity of a noisy channel is defined by
\begin{align}
    u(\varepsilon)=\frac{d}{d-1}\int d\psi \tr[\varepsilon(|\psi\rangle\langle\psi|-\frac{1}{d}\mathbb{I_d})]^2,\notag
    \label{equ_def_unitarity}
\end{align}
where the normalization factor is chosen such that $u(\mathbb{I})=1$.
This way, we define the measuring purity as $p(\rho)=\frac{d}{d-1}\tr(\rho^2)-\frac{1}{d-1}$, which would characterize the incoherent error better than the original purity $p(\rho)=\tr(\rho^2)$. We use a similar fitting function $P_{\text{seq}}=A+Bu^{m-1}$ to fit the decaying of the measuring purity (See Supplemental Information). To compare it directly with each decaying line from RB and plot them together as Fig. \ref{rb}, we draw the incoherent error for each line $1-\epsilon_{\text{inc}}(i)=1-(1-\sqrt{u_i})(d-1)/{d}$, where $i$ can be the reference or the interleaved line. The average purity $P$ for $H_\textrm{L}$ shown in Fig. \ref{rb} is calculated with $P=1-(1-\sqrt{u_{\text{int}}/u_\text{ref}})({d-1})/{d}$.

\bibliographystyle{naturemag}
\bibliography{reference.bib}



\section*{Acknowledgements}
The authors thank Ling-Yan Hung for helpful comments on the manuscript. This work is supported by NSFC grant No. 11875109, 12075110, 11875159, the National Key Research and Development Program of China (No. 2019YFA0308100), General Program of Science and Technology of Shanghai No. 21ZR1406700, Fudan University Original Project (Grant No. IDH1512092/009), Shanghai Municipal Science and Technology Major Project (Grant No. 2019SHZDZX01), Guangdong Innovative and Entrepreneurial Research Team Program (2019ZT08C044), and Science, Technology and Innovation Commission of Shenzhen Municipality (KQTD20190929173815000 and JCYJ20200109140803865). YW is grateful to the Hospitality of the Perimeter Institute during his visit, where the main part of this work is done.

\section*{Author Contributions}
D.L. and Y.W. initiated this work. D.L. supervised the experiments. Y.L., Y.H., and Y.W. elaborated the theoretical framework. Y.F. wrote the computer code and accomplished the NMR experiments. All authors analyzed the data, discussed the results. Y.L, Y.F., Y.W., D.L., and Y.H. wrote the manuscript. Y.F., Y.L. and Y.H. contributed equally to this work.

\section*{Competing Interests}
The authors declare no competing interests.

\section*{Additional Information}

\end{document}